\documentclass{aa}  

\usepackage{graphicx}
\usepackage{txfonts}
\newcommand{\alr}{\mbox{$^{26}$Al}}

\newcommand{\msun}{\mbox{M$_{\odot}$}}
\newcommand{\kms}{\mbox{km~s$^{-1}$}}

\begin{document}

   \title{Observational constraints on the likelihood of $^{26}$Al in planet-forming environments}


   \author{Megan Reiter
          \inst{1}\fnmsep\thanks{megan.reiter@stfc.ac.uk}
          }

   \institute{$^1$UK Astronomy Technology Centre, Blackford Hill, Edinburgh, EH9 3HJ, UK
             }

   \date{Received September 15, 1996; accepted March 16, 1997}

 
  \abstract
  {Recent work suggests that \alr\ may determine the water budget in terrestrial exoplanets as its radioactive decay dehydrates planetesimals leading to rockier compositions. Here I consider the observed distribution of \alr\ in the Galaxy and typical star-forming environments to estimate the likelihood of \alr\ enrichment during planet formation. 
I do not assume Solar-System-specific constraints as I am interested in enrichment for exoplanets generally. 
Observations indicate that high-mass stars dominate the production of \alr\ with nearly equal contributions from their winds and supernovae. 
\alr\ abundances are comparable to those in the early Solar System in the high-mass star-forming regions where most stars (and thereby most planets) form. 
These high abundances appear to be maintained for a few Myr, much longer than the 0.7~Myr half-life. 
Observed bulk \alr\ velocities are an order of magnitude slower than expected from winds and supernovae. 
These observations are at odds with typical model assumptions that \alr\ is provided instantaneously by high velocity mass loss from supernovae and winds.  
Regular replenishment of \alr\, especially when coupled with the small age differences that are common in high-mass star-forming complexes, may significantly increase the number of star/planet forming systems exposed to \alr. 
Exposure does not imply enrichment, but the order of magnitude slower velocity of \alr\ may alter the fraction that is incorporated into planet-forming material. 
Together, this suggests that the conditions for rocky planet formation are not rare, nor are they ubiquitous, as small regions like Taurus that lack high-mass stars to produce \alr\ may be less likely to form rocky planets. I conclude with suggested directions for future studies. }


   \maketitle
%

\section{Introduction}

Recent work from \citet{lichtenberg2019} suggests that the bulk water fraction and radius of terrestrial exoplanets are anti-correlated with the abundance of \alr. 
Heat released by radioactive decay of \alr\ (half-life $\approx 0.72$~Myr) provides an additional heat source that aids planetesimal differentiation \citep[e.g.,][]{hevey2006} and dehydrates planetesimals \citep[e.g.,][]{grimm1993}
leading to a rockier composition of the final planet.

Whether rocky planets are more conducive to life and detectable  bio-signatures is an active area of research \citep[e.g.,][]{kaltenegger2017}. 
There is particular interest in whether water worlds -- those with a significantly higher bulk water fraction than Earth -- can support life \citep[e.g.,][]{noack2016,kite2018,olson2020}. 
Until observations clarify terrestrial exoplanet compositions \citep[e.g.,][]{teske2018,rice2019}, models provide the best constraints on the most likely outcomes of planet formation.

The short half-life of \alr\ 
requires production near planet formation in both time and space. 
Meteoritic evidence indicates that \alr\ was abundant in the early Solar System and most enrichment models aim to reproduce this specific case. 
Early models considered direct supernova (SN) injection into the star-forming cloud, perhaps stimulating its collapse \citep[e.g.,][]{cameron1977} and  
direct enrichment of planet-forming disks \citep[e.g.,][]{chevalier2000}. 
Recently attention has turned to pre-SN mass loss, particularly from Wolf--Rayet (W--R) stars, as a pathway to earlier enrichment \citep[e.g.,][]{gounelle2008,gaidos2009,dwarkadas2017}. 
Scenarios involving either SNe or W--R winds require finely-tuned conditions to place young planet-forming systems in close proximity, suggesting that few systems will be formed under their influence.

Extrapolating from the Solar System has led some authors to suggest that the Earth represents a minority ($\sim$1\%) of the outcomes of planet formation \citep[see e.g.,][]{adams2010,gounelle2015,portegies_zwart2019}. 
However, the specific conditions inferred for the Solar System may not apply to exoplanets in general. 
Abundant \alr\ in many star/planet-forming systems would challenge suggestions that water-rich planets may be the dominant family of terrestrial planet analogues \citep[e.g.,][]{alibert2017,miguel2020}. 
The importance of \alr\ in planet composition clearly warrants a broader assessment of the likelihood of \alr-enrichment.

In this letter, I compare observations of \alr\ in the Galaxy with typical star/planet-forming conditions to estimate the likelihood of \alr\ enrichment.
I focus on \alr\ because of its critical role in determining the water budget of terrestrial exoplanets. My goal is not to explain the formation of the Earth, so I do not consider other elemental abundances (i.e.\ $^{60}$Fe) specific to the Solar System. 
I conclude with a discussion of how models of \alr-enrichment can be improved in light of recent observations.

\section{Observational constraints}\label{s:abundant}

\subsection{Observations of \alr\ and its production by high-mass stars}\label{s:abundant}

Most of our knowledge of the production and distribution of \alr\ in the Galaxy comes from observations of the 1.809~MeV $\gamma$-ray photons produced (along with positrons) when it undergoes radioactive decay \citep{endt1990}. 
Emission is bright along the Galactic plane corresponding to an \textit{average} abundance $\sim$3-25 lower than the early Solar System \citep{lugaro2018}. 
However, the emission 
is predominately associated with young high-mass star-forming regions \citep[e.g.][]{knoedlseder1999_SF26Al,diehl2006}. 
An observed close correlation between 1.809~MeV emission and ionized gas in the interstellar medium (ISM) can only be explained if high-mass stars (>10~\msun) dominate \alr\ production \citep{knoedlseder1999_SF26Al}.

Classical W--R stars are the late evolutionary phase of high-mass stars ($M_{\rm initial}$\, $\gtrsim$\,25\,\msun\ in the Galaxy), and multiple studies argue that their winds contribute $\gtrsim$40\% of the observed \alr\ \citep[e.g.][]{knoedlseder1999_SF26Al}. 
This enhances the ISM abundance of \alr\ prior to the onset of supernovae (SNe). 
\citet{diehl2010} propose a time-evolution of the \alr\ abundance for a given star-forming event as follows:  
\alr\ is initially enhanced in the ISM after 
$\sim$3\,Myr by the onset of mass loss from W--R winds.  
Further enhancement comes over the next 1-2\,Myr from core-collapse SNe, 
with the peak value at $\sim$5\,Myr. 
The abundance then decreases as lower-mass
stars do not reach a W--R phase, making their most significant contribution later from their core-collapse SNe explosions.  Beyond 7-8\,Myr the
\alr\ abundance continues to decrease, with a tail extending to $\sim$20\,Myr.

Population-synthesis models also point to significant early, pre-supernova enrichment \citep[e.g.][]{voss2009}. 
\citet{voss2010,voss2012} modelled the combined high-mass star content of the Orion and Carina star-forming complexes and argued that the current high observed abundance of \alr\ has persisted for $\gtrsim$\,5\,Myr and will be maintained for another $\sim$5\,Myr with additional enhancement from SNe.

Doppler shifts of the 1.809~MeV line indicate that \alr-producing sources corotate with the Galaxy \citep{diehl2006}. 
However, the bulk velocity of \alr-enriched material is higher than that of the molecular gas in the Galaxy, with a systematic offset of $\sim$200\,\kms\ in the direction of Galactic rotation \citep{kretschmer2013}. 
In the specific case of Sco~OB2, \alr\ was detected with slightly blueshifted energies, suggesting a bulk velocity of 137\,$\pm$\,75\,\kms\ \citep{diehl2010}. 
These velocities are roughly an order of magnitude slower than the expected ejection velocities ($>$1000\,\kms) for winds and SNe.

\subsection{Typical star-forming conditions}\label{s:SFeco}

Observations in the Milky Way and in other galaxies suggest a cluster mass function of the form $dN/dM \sim M^{-2}$ \citep[e.g.][]{mok2020}. 
I refer to clustering in the statistical sense that stars form near other stars \citep[see discussion in][]{lee2020}, noting that many extra-galactic studies do not apply additional criteria to determine whether stellar over-densities are gravitationally bound.
As \citet{dukes2012} point out, this cluster mass function implies that $\gtrsim$50\% of stars are born in high-mass star-forming regions with as many or more high-mass stars than the Orion Nebula Cluster (ONC).

The Orion star-forming complex is nevertheless observed to have \alr\ \citep{diehl2003} at a level similar to the early Solar System \citep{jura2013}. 
The same is true of Carina \citep{reiter2019}. 
\citet{jura2013} also made a more general argument by comparing the Galactic \alr\ abundance with molecular (star-forming) gas to show that star/planet formation regions have abundances on the order of the early Solar System, unlike the average ISM.


\section{Discussion}

The majority of stars ($\gtrsim 50$\%) form in high-mass regions where \alr\ is observed to be abundant. 
While this suggests that many systems are exposed to \alr, two key uncertainties affect the fraction of systems that are ultimately enriched: 
(1) when and how long \alr\ is highly abundant  
and 
(2) how and whether the \alr\ mixes with the star/planet forming material.
I discuss these in turn below.

\subsection{The timescale of \alr\ availability}\label{ss:availability}

Observations of \alr\ indicate a fundamental discrepancy with typical model assumptions: high \alr\ abundances appear to persist for a few Myr, much longer than the 0.7\,Myr half-life assumed for an instantaneous production event.
The longer timescale reflects production by a stellar population rather than a single high-mass source.

For models that assume a single-age stellar population, disks must survive long enough for direct pollution from a nearby SN or W--R star. Prior to these late evolutionary stages, 
feedback from those same high-mass stars may accelerate disk destruction, limiting the number of remaining systems to enrich \citep[e.g.,][]{winter2018,concha2019}. 

Other models start with a SN or W--R winds polluting nearby cold gas that will subsequently form a second (or third) generation of stars \citep[e.g.,][]{gaidos2009,gounelle2012,kuffmeier2016,dwarkadas2017}. 
However, several studies show that most of the gas is cleared from star-forming regions by pre-SN feedback \citep[e.g.][]{hollyhead2015,chevance2019}.

Reality likely lies between these scenarios. 
Observations consistently suggest small age spreads in star-forming regions \citep[e.g.][]{getman2014,beccari2017,venuti2018,jerabkova2019,prisinzano2019}. 
Age differences -- where younger and older populations are not necessarily spatially coincident -- are also common within OB associations \citep[e.g.,][]{wright2010,pecaut2016,getman2018}. 
In both cases, age differences are a few Myr, much smaller than would be produced by triggered sequential star formation \citep{parker_dale_2016}.

Age differences affect \alr-enrichment in two ways: 
(1) they extend the time disks are in the region \citep[as in the proplyd lifetime solution proposed by][]{winter2019_proplyds}; and (2) they provide regular replenishment to maintain the high abundance of \alr\ for several Myr. 
This pragmatic approach is more akin to the hierarchical star formation models from \citet{fujimoto2018} that show galactic-scale correlations in star formation strongly affect enrichment.

Take as an example, the Carina star-forming complex. 
Several studies suggest age differences between the primary clusters Tr14, Tr15, and Tr16 (see, e.g., the discussion in \citealt{townsley2011} for a review). 
Recent work from \citet{povich2019} quantifies variations in the duration of star formation 
in Carina. 
They conclude 
that star formation has been ongoing for $\sim$10~Myr, with a peak in the star-formation rate $\sim$3~Myr ago with the birth of the famous central clusters, Tr14 and Tr16.

These multiple episodes of star formation are all contained within the $\gtrsim$\,1$^{\circ}$ beamsize of the $\gamma$-ray observations.  
\citet{voss2012} performed population synthesis modelling of the combined population of all clusters to estimate that the current high abundance has been maintained for $\sim$5~Myr and will persist for another $\sim$5~Myr.

If \alr\ abundances have been maintained at the current high levels for $\sim$5~Myr while star formation peaked within the last $\sim$3~Myr, then the majority of systems were likely exposed to \alr. 
Similar arguments can be made for the Orion star-forming complex. 
If $\gtrsim 50$\% of all stars are born in high-mass regions with (conservatively) half of those exposed to significant \alr, then we may crudely estimate that as many as $\sim$25\% of systems may be enriched with \alr, possibly at a level adequate to make rocky planets.  
This is an order of magnitude larger than the few percent estimated for Solar-System-specific conditions, but consistent with model results from \citet{fujimoto2018} that short-lived radioactive isotopes like \alr\ are abundant in newly formed stars.

This order-of-magnitude estimate does not account for 
uncertainties in the cluster mass function and ignores possible variations with location in a galaxy. 
No provision is made for differences in \alr\ production with stellar mass, although not all high-mass stars undergo a W--R phase and SN yields depend on the mass of the progenitor. 
More quantitative analysis is clearly needed. I outline some suggested directions in Section~\ref{s:suggestions}.

\subsection{The incorporation of \alr\ into planet-forming material}\label{ss:incorp}

Exposure to \alr\ does not imply that it mixes with planet-forming material. 
Here too observations invite reconsideration of typical model assumptions.  
The observed bulk velocity of the \alr\ is $\sim$200~\kms,
an order of magnitude slower than the $\sim$1000~\kms\ expected from W--R winds or SNe (see Section~\ref{s:abundant}). 
Slower \alr\ may mix more readily with planet-forming material. 
Existing parameter studies of disk enrichment investigate only a narrow range of velocities that are much faster than the observed bulk \alr\ velocity \citep[e.g.,][]{ouellette2007}.

Other models for the enrichment of star/planet-forming gas enforce mixing by confining \alr-enriched ejecta \citep[e.g.,][]{vasileiadis2013,gounelle2015,kuffmeier2016}. 
This would preclude \alr-enrichment in regions like Carina that have developed into superbubbles unable to contain \alr-enriched material. As pointed out by \citet{fujimoto2018}, these scenarios are at odds with both the short observed time for young high-mass star-forming regions to clear their gas and the large observed scale height of \alr\ in the Galaxy \citep{wang2009}.

While the physical pathway by which \alr\ is incorporated into planet-forming material remains unclear, composition studies of extrasolar asteroids suggests that mixing nevertheless does occur. 
\citet{jura2013} estimate the \alr\ required to enable the observed differentiation and find that it is comparable to the \alr\ abundance in Orion, implying that Solar-System-levels of enrichment are common. 

Additional evidence may come from the observed size distribution of exoplanets.  
The models of \citet{lichtenberg2019} predict that the planet radius is anti-correlated with the bulk water fraction.  
Observations of small, close-in exoplanets represent a bimodal radius distribution 
with current data suggesting the same intrinsic frequency for the two regimes \citep{fulton2017}.

\subsection{Some environments are more likely to form rocky planets}\label{ss:location}

The points in the previous sections suggest that some star-forming environments will be more conducive than others to \alr-enrichment, and thus rocky planet formation. 
The most massive member of Taurus, the local template of a low-mass star-forming region, is $\sim$3.5\,\msun\ \citep[Table~1,][]{demarchi2010}. Such a star will end its life as an AGB star that will not sustain temperatures sufficient to synthesize \alr\ \citep[e.g.,][]{abia2017}. 
If \alr\ regulates the water budget, as in the models of \citet{lichtenberg2019}, this suggests that terrestrial planets that form in low-mass regions like Taurus are more likely to be water worlds. 


\subsection{Future directions}\label{s:suggestions}

In the following sections, I highlight key areas for future work to quantify the fraction of systems that are enriched with \alr\ and therefore able to form rocky planets.

\subsubsection{Production of \alr\ by high-mass stars}\label{ss:HMmodels}

Stellar mass is only one of the relevant variables affecting the timescale and abundance of \alr\ production. 
Rotation is a key parameter that influences the amount of \alr\ introduced
into the ISM by high-mass stars, primarily via rotational-mixing of material from 
deeper in the star to the surface \citep[e.g.][]{voss2009}. By bringing \alr\ to the
surface earlier than in models without rotation, stars can potentially return \alr\ to the ISM via their stellar winds both earlier and for lower initial masses than in non-rotating models. 
For 
\alr-enrichment the effects of rotation have been explored for initial rotational velocities of 300\,\kms\ \citep[e.g.][]{palacios2005,voss2009}.

Surveys such as the VLT-FLAMES Tarantula Survey \citep{cjevans2011} and the IACOB project \citep{ssd11a,ssd11b,ssd15} have provided
large spectroscopic samples to investigate the rotation rates of high-mass stars
\citep[e.g.][]{ramirez2013,holgado2018,holgado19} 
Typical rotation rates are less than half of those assumed in population synthesis studies \citep[e.g.,][]{voss2009}. 
At these velocities, stellar evolution more closely resembles non-rotating models \citep{evans2020}. 
Existing population synthesis models matched the observed \alr, so it is unclear if changes to \alr\ production also affect the time-evolution of the abundance.

Another critical development in the last 10 years is evidence that $>70$\% of high-mass stars are in close-separation binaries that will eventually interact \citep{sana2012}. 
Large grids of binary models 
are only now being computed.
Recent work from \citet{brinkman2019} presents non-rotating models that suggest that binary interactions affect the total \alr\ mass ejected by only $\sim$5-10\%. There is a substantial enhancement of \alr\ from 10-15\,\msun\ primaries in binary systems, but the total amount from a given population is dominated by higher mass stars ($>$30\,\msun), where binary effects appear to be less important. Further work is required to investigate the impact of stellar rotation on such models.

\subsubsection{Pre-supernova enrichment models}\label{ss:enrichment_models}

Studies of the $\gamma$-ray emission from the radioactive decay of \alr\ point to the important role of pre-SN mass loss. 
Including pre-SN enrichment in cluster-scale simulations \citep[e.g.][] {lichtenberg2016,nicholson2019} may significantly change the fraction of disks enriched with \alr. Models should also consider on-going \alr\ production, and perhaps steady accretion instead of a single injection event. Meteoritic evidence suggests this may have happened in the Solar System with \alr\ accrued quickly, but not instantaneously \citep{liu2012}.  

\section{Summary}

Recent models suggest that the \alr\ abundance plays a central role in determining the water budget of terrestrial planets. 
Galactic observations of the short-lived radioactive isotope \alr\ provide a constraint on the fraction of star/planet-forming systems that may be enriched. 
\alr\ is predominantly produced
by high-mass stars. Pre-SN mass loss, especially from W--R stars, is thought to contribute nearly half the Galactic \alr\ budget. \alr\ is particularly abundant in high-mass star-forming regions where the majority of stars, and thus planets, form. Mass loss by stellar winds can enrich the ISM in such regions in a few Myr. 
Critically for planet formation, regular replenishment may sustain high levels of \alr\ for Myrs (much longer than its 0.7\,Myr half-life). 
I argue that the majority of star/planet systems in high-mass star-forming regions may be exposed to abundant \alr. 
Exposure does not imply enrichment, but observed bulk \alr\ velocities are an order of magnitude slower than expected from winds and SNe, which may increase the likelihood of enrichment.
If half of the exposed systems are enriched with \alr, this (rough) estimate suggests that the fraction of such systems is at least an order of magnitude higher than the few percent found by studies that extrapolate from Solar-System-specific conditions. 
More quantitative estimates that do not assume Solar-System-specific conditions a priori are clearly needed to clarify this fraction.

These conditions describe the environment in which most stars and planets form, but they do not reflect those sampled in low-mass star-forming clouds. 
This suggests that rocky planets are more likely to form around stars born in high-mass star-forming regions. 
Low-mass regions such as Taurus do not contain the high-mass stars that dominate the production of \alr\ and thus are unlikely to form rocky planets.

\begin{acknowledgements}
      I would like to thank the referee for a timely and thoughtful report that improved the manuscript. I am deeply grateful to Chris Evans, Richard Parker, Ken Rice, and Tim Lichtenberg for reading the manuscript and providing thoughtful comments. Thanks also to Brandt Gaches and Karen Moran. This project received funding from the European Union’s Horizon 2020 research and innovation programme under the Marie Sk\'{l}odowska-Curie grant agreement No.\ 665593 awarded to the Science and Technology Facilities Council.
\end{acknowledgements}

%
   \bibliographystyle{aa} 
   \bibliography{bibliography_mrr} 
%

\end{document}